\documentclass[letterpaper, 10 pt, conference]{ieeeconf} 
\IEEEoverridecommandlockouts
\usepackage[english]{babel}
\usepackage{cite}
\usepackage{amsmath,amssymb,amsfonts}
\usepackage{algorithmic}
\usepackage{graphicx}
\usepackage{subcaption}
\usepackage{textcomp}
\usepackage{amsmath}
\usepackage{xcolor}
\usepackage{units}
\usepackage{booktabs}
\usepackage{comment}
\usepackage[hidelinks]{hyperref}
\usepackage[normalem]{ulem}
\usepackage{float}
\usepackage{url}
\usepackage{dsfont}
\usepackage{titlesec}
\newtheorem{thm}{Theorem}
\newtheorem{prob}[thm]{Problem} 
\titlespacing*{\section}{0pt}{0.9pt}{0.9pt}
\titlespacing*{\subsection}{0pt}{0.5pt}{0.5pt}
\title{\LARGE \bf Stochastic Model Predictive Control of Charging Energy Hubs with Conformal Prediction} 

\author{Diego Fernández-Zapico, Theo Hofman, Mauro Salazar
\thanks{Control Systems Technology section, Eindhoven University of Technology, Eindhoven, The Netherlands
        {\tt\small \{d.fernandez.zapico, t.hofman, m.r.u.salazar\}@tue.nl}}%
}

\begin{document}
\maketitle
\begin{abstract}                
This paper presents an online energy management system for an energy hub where electric vehicles are charged combining on-site photovoltaic generation and battery energy storage with the power grid, with the objective to decide on the battery (dis)charging to minimize the costs of operation.
To this end, we devise a scenario-based stochastic model predictive control (MPC) scheme that leverages probabilistic 24-hour-ahead forecasts of charging load, solar generation and day-ahead electricity prices to achieve a cost-optimal operation of the energy hub.
The probabilistic forecasts leverage conformal prediction providing calibrated distribution-free confidence intervals starting from a machine learning model that generates no uncertainty quantification.
We showcase our controller by running it over a 280-day evaluation in a closed-loop simulated environment to compare the observed cost of two scenario-based MPCs with two deterministic alternatives: a version with point forecast and a version with perfect forecast. Our results indicate that, compared to the perfect forecast implementation, our proposed scenario-based MPCs are 13\% more expensive, and 1\% better than their deterministic point-forecast counterpart.
\end{abstract}


\section{Introduction}
The rapid electrification of the transport sector requires an infrastructure capable of charging large fleets of electric vehicles (EV).  The European Parliament and The Council  approved the alternative fuels infrastructure regulation, setting a minimum of one off-road charging station every 60 km by 2025, for cars and vans on main European transport corridors \cite{EU2023}.
Furthermore, the Government of the Netherlands  presents the Climate Agreement with the goal for all new passenger vehicles to be emission-free by 2030 \cite{NCA2019}. Energy hubs can support EV charging stations to meet the increased demand by using renewable energy sources and a battery energy storage system (BESS).

The energy management system (EMS) used to operate energy hubs often employs rule-based methods \cite{ChandraSinghEtAl2022} and heuristics. In \cite{EngelhardtZepterEtAL2022} authors propose a heuristic-based control of a grid inverter for an energy hub composed of a photovoltaic (PV) installation, a multi-battery BESS and EV chargers. As inputs the BESS state of charge and the one-hour ahead PV forecast are employed, neglecting key factors such as EV charging demand or the cost of electricity exchanged with the grid. Additionally, it is limited by a PV forecast with a short prediction horizon and no measure of uncertainty. In \cite{YanZhangEtAl2019} authors overcome these limitations with a multi-horizon chance-constrained optimization control for day-ahead, hour-ahead and real-time operation. However, it relies on distributional assumptions for uncertain variables, limiting its applicability. In \cite{ZhangSunEtAl2023} authors propose a model predictive control (MPC) that solves a second-order cone optimization problem at every time step, using a scenario-based approach to account for the uncertainty of charging demand and solar generation in the energy hub. However, the scenarios are obtained by assuming a distribution of the historical data.

Recently, a systematic review of day-ahead EV load forecasting shows the best performing models (extreme gradient boosting, multi-layer perceptron) make no such assumptions about the distribution of the data \cite{BamposLaitsosEtAl2024}. In a literature review of day-ahead electricity price forecasting, authors introduce best practices and show that the best-performing model is a deep neural network \cite{LagoMarcjaszEtAl2021}, which does not rely on distributional assumptions. In recent tutorial on PV power forecasting, authors include gradient-boosted models among the best performers, while emphasizing the importance of feature engineering \cite{YangXiaEtAl2024}. Starting from the state-of-the-art machine learning and deep learning models mentioned above, probabilistic forecasts can be obtained using conformal prediction (CP). In \cite{XuXie2021} authors introduce EnbPI, an algorithm based on CP for distribution-free uncertainty quantification of any forecasting model. 

The performance of the EMS of an energy hub directly depends on the forecasting accuracy of the predicted variables. To the best of our knowledge, there is no published work on EMS leveraging state-of-the-art forecasting models with calibrated distribution-free uncertainty intervals using CP. Therefore, we contribute by introducing a probabilistic prediction module for 24-hour-ahead EV load, PV generation and day-ahead electricity prices based on Gradient-boosted Trees \cite{Friedman2001} and EnbPI. Then we evaluate the accuracy and coverage of the predictions and quantify the operational cost of two scenario-based stochastic MPCs in a closed-loop simulated environment.  

\section{Charging Energy Hub Model}\label{model}
\begin{figure}[t]
	\centering
	\includegraphics[width=0.63\linewidth]
    {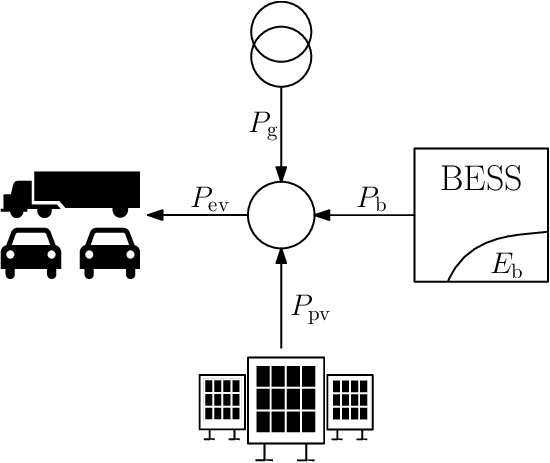}
    \vspace{-4pt}
	\caption{
     Power flow of the Charging Energy Hub.}
	\label{fig:ceh}
\end{figure}

In Fig. \ref{fig:ceh}, we present the power flow of the energy hub considered.
It is composed of a grid connection ($P_\mathrm{g}$), a BESS ($P_\mathrm{b}$), a solar plant ($P_\mathrm{pv}$) and the total EV charging demand ($P_\mathrm{ev}$). We work in discretized time with index $k$ and time resolution $\Delta T$. At each step $k$ and for each scenario $s$, the power balance of the energy hub is satisfied
\begin{equation}\label{eq:Pbalance}
    P_{\mathrm{ev},k}^s = P_{\mathrm{g},k}^s + P_{\mathrm{pv},k}^s + P_{\mathrm{b},k}^s,
\end{equation} 
where we use $s$ to describe variable uncertainties and each model component is introduced in the remainder of this section.

\subsection{BESS}
We model the BESS following the methodology in \cite{ClementeSalazarEtAl2024}, which models the losses of the battery quadratically
\begin{equation}\label{eq:BlossOriginal}
    P_{\mathrm{ib}, k}^s = P_{\mathrm{b}, k}^s + \frac{(P_{\mathrm{ib}, k}^s)^2}{P_{\mathrm{sc}, k}^s}, 
\end{equation}
and relaxes it as a second-order conic constraint
\begin{equation}\label{eq:Bloss1}
    P_{\mathrm{ib}, k}^s - P_{\mathrm{b}, k}^s + P_{\mathrm{sc}, k}^s  \geq \left\| \begin{matrix} 2 P_{\mathrm{ib}, k}^s  \\ P_{\mathrm{ib}, k}^s - P_{\mathrm{b}, k}^s - P_{\mathrm{sc}, k}^s  \end{matrix} \right\|_2,
\end{equation}
where $P_{\mathrm{ib}, k}^s$ is the internal battery power for scenario $s$ and at step $k$, $P_{\mathrm{b}, k}^s$ is the external battery power and $P_{\mathrm{sc}, k}^s$ is the short-circuit power. This is approximated based on 
\begin{equation} \label{eq:Bloss2}
    P_{\mathrm{sc}, k}^s \leq a_m \cdot E_{\mathrm{b}, k}^s + b_m  \quad \forall m \in \mathcal{M}, 
\end{equation}
where $E_{\mathrm{b}, k}^s$ is the battery capacity and $a_m$ and $b_m$ are the coefficients of a piecewise linear of each spline $m$ in the set of all splines $\mathcal{M}$, which are obtained from a fit of measures on $P_\mathrm{sc}$ and $E_\mathrm{b}$. The dynamics of the battery are a discrete linear expression
\begin{equation}\label{eq:EbDyn}
    E_{\mathrm{b},k+1}^s = E_{\mathrm{b},k}^s - \Delta T \cdot P_{\mathrm{ib}, k}^s.
\end{equation}
Additionally, the battery is constrained to operate between the minimum ($\underline{E}_{\mathrm{b}}$) and maximum ($\overline{E}_{\mathrm{b}}$) operational capacity
\begin{equation}\label{eq:Eblims}
    E_{\mathrm{b},k}^s \in [\underline{E}_{\mathrm{b}}, \overline{E}_{\mathrm{b}}].
\end{equation}

\subsection{Total EV Charging Demand}
We model the power demand of charging EVs in an aggregated manner. We assume each EV charges with an average power profile
\begin{equation}
    \overline{P}_{\mathrm{c},i} = \frac{E_{\mathrm{c},i}}{\Delta T \cdot (k_{f, i} - k_{0,i})},    
\end{equation}
where $E_{\mathrm{c},i}$ is the energy required by EV $i$ between the connection step ($k_{0, i}$) and the disconnection step ($k_{f, i}$). The total EV power demand at $k$ is the sum over all charging EVs 
\begin{equation}
    P_{\mathrm{ev}, k} = \sum_{i=1}^{N_\mathrm{ev}} \overline{P}_{\mathrm{c},i} \cdot \mathds{1}_{i,k},
\end{equation}
where $N_\mathrm{ev}$ is the total number of EVs and $\mathds{1}_{i,k}$ is a binary condition indicating if charging is taking place
\begin{equation}
    \mathds{1}_{i,k} = \begin{cases}
        1  & \text{if } k \in [k_{0, i}, k_{f, i}] \\
        0 & \text{otherwise}.
    \end{cases}
\end{equation}

Finally, for each $s$, we assume that $P_{\mathrm{ev}, k}^s$ is the power measured at the meter, after discounting for component losses.

\subsection{Power Grid}
We consider the charging station can buy and sell power from the grid $P_{\mathrm{g}, k}^s$ at the day-ahead price of electricity, incurring an operational cost per unit time $\Delta T$.
\begin{equation}\label{eq:powergridOriginal}
    C_{\mathrm{el}, k}^s = 
    \begin{cases}
    p_{\mathrm{buy},k}^s \cdot P_{\mathrm{g}, k}^s & \text{if } P_{\mathrm{g}, k}^s \ge 0\\ 
    p_{\mathrm{sell}, k}^s \cdot P_{\mathrm{g}, k}^s & \text{otherwise},
    \end{cases}
\end{equation}
where $p_{\mathrm{buy},k}^s$ and $p_{\mathrm{sell},k}^s$ are the buying and selling prices, respectively, which we define in detail in~\eqref{eq:priceQH} and~\eqref{eq:sellbuyratio}. Since we aim to minimize the total cost of operation, assuming $p_{\mathrm{buy},k}^s\geq p_{\mathrm{sell},k}^s$, this constraint can be losslessly relaxed in a convex manner~\cite{BorsboomFahdzyanaEtAl2021} as
\begin{equation}\label{eq:powergrid}
    C_{\mathrm{el}, k}^s \geq 
    \begin{cases}
    p_{\mathrm{buy},k}^s \cdot P_{\mathrm{g}, k}^s\\ 
    p_{\mathrm{sell}, k}^s \cdot P_{\mathrm{g}, k}^s.
    \end{cases}
\end{equation}



\subsection{Solar Plant}
At each $k$ and for each $s$, we assume that the PV power ($P_{\mathrm{pv}, k}^s$) is the power generated by the solar plant measured at the meter level, after discounting for component losses. 

\section{Forecasting}\label{probforecasting}
To control the energy hub model of Section \ref{model}, we need to predict the unknown variables on the prediction horizon required by the MPC. We consider as unknown predictable variables: $P_\mathrm{ev}$, $P_\mathrm{pv}$ and the day-ahead price of electricity ($p_\mathrm{el}$).
In this section, we describe how we compute probabilistic forecasts and scenarios of $P_\mathrm{ev}$, $P_\mathrm{pv}$ and $p_\mathrm{el}$.

\subsection{Probabilistic Forecasting}
Here we introduce how we learn a response variable $Y$ with a model $F$ based on features $x$. We use Gradient-boosted Trees \cite{Friedman2001} 
\begin{equation}\label{eq:f_det}
    F(x) = F_0(x) + \sum_{m=1}^{N_m} \nu \cdot \rho_m  \cdot h_m(x),
\end{equation}
where $F_0$ is the initial prediction, $\nu$ is the learning rate, $\rho_m$ is the line search on each boosting round $m$, $h_m$ are decision trees and $N_m$ are the total number of boosting rounds. We train the model to minimize the absolute error ($|Y - F(x)|$) on the training set, since the cost associated to our problem is linear \cite{LagoMarcjaszEtAl2021}. To avoid overfitting, we use early stopping with a validation set. We use the GradientBoostingRegressor implementation from Scikit-learn \cite{PedregosaVaroquauxEtAl2011}.

Starting from the base model in~\eqref{eq:f_det}, we produce a calibrated distribution-free prediction interval using EnbPI \cite{XuXie2021}
\begin{equation}\label{eq:f_prob}
    \mathrm{PI}_\alpha(x) = \hat{F}^\phi(x) \pm (1-\alpha) \, \mathrm{quantile} (\hat{\epsilon}^\mathrm{LOO}),
\end{equation}
where $\alpha$ is the significance level, $\hat{F}^\phi$ is the mean of the leave-one-out (LOO) estimators obtained from fitting $F$ on each LOO training sample,  $\hat{\epsilon}^\mathrm{LOO}$ is the absolute value of the residuals of the LOO estimators on the corresponding calibration sets ($|Y - \hat{F}^\mathrm{LOO}(x)|$). For this part, we use the EnbPI implementation from MAPIE \cite{CordierBlotEtAl2023}. 

\subsection{Feature Engineering}
At each $k$ and for each scenario $s$, we predict $P_\mathrm{ev}$ and $P_\mathrm{pv}$ on a \unit[24]{h} window with \unit[15]{min} resolution and length $N_\mathrm{wq}=96$,
\begin{equation}
    \hat{\pmb{P}}_{\mathrm{ev},k}^s = [\hat{P}_{\mathrm{ev}, k}^s, ..., \hat{P}_{\mathrm{ev}, k+N_\mathrm{wq} - 1}^s]^{\top} \in \mathbb{R}^{N_\mathrm{wq}},
\end{equation}
\begin{equation}
    \hat{\pmb{P}}_{\mathrm{pv}, k}^s = [\hat{P}_{\mathrm{pv}, k}^s, ..., \hat{P}_{\mathrm{pv}, k+N_\mathrm{wq} - 1}^s]^{\top} \in \mathbb{R}^{N_\mathrm{wq}},
\end{equation}
where  $\hat{P}_{\mathrm{ev},k}^s$ denotes the prediction of $P_{\mathrm{ev},k}$ for scenario $s$.
 Following \cite{YangXiaEtAl2024}, we design $x$ to predict $P_\mathrm{pv}$ using weather forecast information (direct radiation, diffuse radiation, temperature, wind speed), the sun position (zenith angle, solar time), calendar information (month of the year) and lagged daily power features (sum, standard deviation). The one-day lag of the daily total power reads
\begin{equation}\label{eq:laggSum}
    P_{\mathrm{pv}, d-1}^{\mathrm{sum}} = \sum_{k \in \mathcal{T}_{d-1}} P_\mathrm{pv, k},
\end{equation}
where $d$ is the day index ($d = \lfloor \frac{k}{96} \rfloor$) and $\mathcal{T}_{d-l} = \{ k \mid 96\cdot(d-l) \leq k < 96\cdot(d-l+1) \}$. Similarly, we define the one-day lag of the daily standard deviation of the power as
\begin{equation}\label{eq:laggStd}
    P_{\mathrm{pv}, d-1}^{\mathrm{std}} = \sqrt{\frac{1}{96} \cdot \sum_{k \in \mathcal{T}_{d-1}} (P_{\mathrm{pv}, k} - P_{\mathrm{pv}, d-1}^{\mathrm{avg}})^2},
\end{equation}
where $P_{\mathrm{pv}, d-1}^{\mathrm{avg}} = \frac{1}{96} \cdot P_{\mathrm{pv}, d-1}^{\mathrm{sum}}$.

We design the features to predict $P_\mathrm{ev}$ in a similar way. We use calendar information (month of the year, day of the week, arrival hour), lagged daily power features (sum, standard deviation) and a lagged intraday power feature (sum). The lagged daily power features are equivalent to~\eqref{eq:laggSum} and~\eqref{eq:laggStd}. For the intraday feature, we use the sum of the sixth lagged hour
\begin{equation}\label{eq:featEvlagIntra}
    P_{\mathrm{ev}, h-6}^{\mathrm{sum}} = \sum_{k \in \mathcal{T}_{h-6}} P_\mathrm{ev, k},
\end{equation}
where $\mathcal{T}_{h-6} = \{ k \mid k - 28 + k \leq k < k - 24\}$.

We forecast $p_\mathrm{el}$, at each $k$ and for each scenario $s$, on a \unit[24]{h} window with \unit[1]{h} resolution and length $N_\mathrm{w h}=24$
\begin{equation}
    \hat{\pmb{p}}_{\mathrm{el},k}^s = [\hat{p}_{\mathrm{el}, k}^s, ..., \hat{p}_{\mathrm{el}, k+N_\mathrm{wh}-1}^s]^{\top}  \in \mathbb{R}^{N_\mathrm{wh}},
\end{equation}
where we assume there is no day-ahead auction mechanism and, therefore, we treat the price as a real time signal. Furthermore, to capture trends, we work with a \unit[24]{h} difference of the time series
\begin{equation}
    \hat{\pmb{p}}_{\mathrm{el},k}^{\prime s}=
    \hat{\pmb{p}}_{\mathrm{el},k}^s - {\pmb{p}}_{\mathrm{el},k-24},
\end{equation}
where ${\pmb{p}}_{\mathrm{el},k-24} = [p_{\mathrm{el}, k-24}, ...,  p_{\mathrm{el}, k-1}]^{\top}  \in \mathbb{R}^{N_\mathrm{wh}}$ are the observed values required to compute the \unit[24]{h} differences.
We design the features to predict $p_{\mathrm{el}}^\prime$ following \cite{LagoMarcjaszEtAl2021}. We use calendar information (month of the year, day of the week, hour of the day), a lagged daily price feature (standard deviation) and two exogenous forecasts: national electricity demand ($P_\mathrm{load}$) and national solar and wind generation ($P_\mathrm{gen}$). The lagged daily price feature is equivalent to~\eqref{eq:laggStd}. For the features with exogenous forecasts of $P_\mathrm{load}$ and $P_\mathrm{gen}$, we work with normalized \unit[24]{h} differences,
\begin{equation}
	\hat{P}_{\mathrm{exog},k}^{\prime}=
	\frac{\hat{P}_{\mathrm{exog},k} - \hat{P}_{\mathrm{exog},k-24}}{P_{\mathrm{exog}, d}^\mathrm{max}},
\end{equation}
where $P_{\mathrm{exog}, d}^\mathrm{max} = \max (\{ P_{\mathrm{exog}, k} \}_{k \in \mathcal{T}_d} )$ and, for conciseness, subscript $\mathrm{exog}$ refers to either of the two exogenous forecasts.
    We assume the exogenous forecasts are available  at every $k$ for all hours of the predicted window. To obtain the buying price required in~\eqref{eq:powergrid}, we convert the predicted day-ahead electricity prices as,
    \begin{equation}\label{eq:priceQH}
        \hat{p}_{\mathrm{buy},k + h + q}^s = \Delta T \cdot \hat{p}_{\mathrm{el},k + h}^s \quad \forall h \in \mathcal{T_H} ,  q \in \mathcal{T_Q},
    \end{equation}
    where $h$ is the hour index, the hour set $\mathcal{T_H} = \{0, ..., 23\}$, $q$ is the quarter hour index, the quarter hour set $\mathcal{T_Q} = \{ 0, 1, 2, 3\}$ and $\Delta T  = \unit[0.25]{h}$. Additionally, we define the selling price from~\eqref{eq:powergrid} as
    \begin{equation}\label{eq:sellbuyratio}
        \hat{\pmb{p}}_{\mathrm{sell},k}^s = \hat{\pmb{p}}_{\mathrm{buy},k}^s.
    \end{equation}
    where at each step $k$ and for each scenario $s$, the resulting vectors $ \hat{\pmb{p}}_{\mathrm{sell},k}^s ,  \hat{\pmb{p}}_{\mathrm{buy},k}^s \in \mathbb{R}^{N_\mathrm{wq}}$. More details on this section are available on the public repository of the article \footnote{\url{{https://github.com/diegofz/ChargingEnergyHubs_MPC}}}.

\subsection{Generating Scenarios}\label{gen_scenarios}
For each predicted variable, we select $n_s = 1 + n_\alpha$ scenarios from~\eqref{eq:f_prob}, where $n_\alpha$ is the number of percentiles used. Then we build a stochastic tree with the $n_s$ scenarios from each variable, assuming independence between variables. The resulting stochastic tree has $N_s = n_s ^ {n_v}$ branches, where $n_v$ is the number of variables considered. The probability of each branch is $\rho_s$. For our case, we build a stochastic tree for the predicted variables ($n_v=3$) and select $n_s=3$ scenarios from~\eqref{eq:f_prob}: $\hat{F}^\phi$, the 5$^\mathrm{th}$ and the 95$^\mathrm{th}$ percentiles. 
    
\section{Model Predictive Control}
Based on the scenarios generated from the predictions of $P_\mathrm{ev}$, $P_\mathrm{pv}$ and $p_\mathrm{el}$ in Section \ref{probforecasting}, we use a scenario-based stochastic MPC to minimize costs of operation of the energy hub. We aim to control daily evaluation episodes with length $N_\mathrm{ep}=96$ and time step set $\mathcal{K} = \{ 0, ..., N_\mathrm{ep} - 1\}$. We use $P_{\mathrm{ib}}$ as the control variable which produces the transition defined in~\eqref{eq:EbDyn} on the state variable $E_\mathrm{b}$. We introduce a periodic constraint on the state variable
\begin{equation}\label{eq:periodicity}
    E_{\mathrm{b}, 0}^s = E_{\mathrm{b}, N_\mathrm{ep}}^s,
\end{equation}
which ensures a fair comparison between  different evaluation episodes, since all episodes start and end with the same state. We also constrain the control variable to be the same across all scenarios
\begin{equation}\label{eq:scenario}
    P_{\mathrm{ib}, i|k}^{1} = P_{\mathrm{ib}, i|k}^{s} \quad \forall s \in \mathcal{S}, i \in \mathcal{R},
\end{equation}
where $\mathcal{S}=\{1, ..., N_s\}$, $i$ is the time step within the receding window of length $N_\mathrm{w}=96$ and we use notation  $i|k = k+i$ to indicate that step $i$ is conditioned on information available at step $k$. We define $\mathcal{R}$ depending on the version of the MPC. For the Stochastic MPC we set $\mathcal{R}=\{0, ..., N_\mathrm{w}-1\}$ and for the Recourse MPC, we relax this constraint by setting $\mathcal{R}=\{0\}$.

At each $k$ and for each $s$, we compute predictions $\hat{\pmb{P}}_{\mathrm{ev}, k}^s$, $\hat{\pmb{P}}_{\mathrm{pv}, k}^s$, $\hat{\pmb{p}}_{\mathrm{buy}, k}^s$, $\hat{\pmb{p}}_{\mathrm{sell}, k}^s$ and initialize the variables
\begin{equation}
    P_{\mathrm{ev}, i|k}^s = \hat{P}_{\mathrm{ev}, i}^s \quad \forall i \in \mathcal{I},
\end{equation}
\begin{equation}
    P_{\mathrm{pv}, i|k}^s = \hat{P}_{\mathrm{pv}, i}^s \quad \forall i \in \mathcal{I},
\end{equation}
\begin{equation}
    p_{\mathrm{buy}, i|k}^s = \hat{p}_{\mathrm{buy}, i}^s \quad \forall i \in \mathcal{I},
\end{equation}
\begin{equation}
    p_{\mathrm{sell}, i|k}^s = \hat{p}_{\mathrm{sell}, i}^s \quad \forall i \in \mathcal{I},
\end{equation}
where $\mathcal{I}=\{0, ..., N_\mathrm{w}-1\}$. Similarly, we convert the time index $k$ in~\eqref{eq:Pbalance},~\eqref{eq:Bloss1},~\eqref{eq:Bloss2},~\eqref{eq:EbDyn},~\eqref{eq:Eblims} and ~\eqref{eq:powergrid} to $i|k$ $\forall i \in \mathcal{I}$.
We then solve the second-order cone program in Problem~\eqref{prob:mpcSce} to obtain the cost-optimal control and state trajectories. As the objective function, we use the average cost of operation across the scenarios
\begin{equation}
    J_{i|k} = \sum_{s=1}^{N_s}\rho_s \cdot C_{\mathrm{el}, i|k}^{s},
\end{equation}
where we assume equally probable scenarios ($\rho_s = \frac{1}{N_s}$).

\begin{prob}[Scenario-based Model Predictive 
Control]\label{prob:mpcSce}
\begin{equation*}
\begin{aligned}
&\!\min _{\pmb{P}_{\mathrm{ib}, k}} & & \sum_{i=0}^{N_\mathrm{w}-1} J_{i|k} \\
& \textnormal{s.t.} & &  (\ref{eq:Pbalance}), (\ref{eq:Bloss1}), (\ref{eq:Bloss2}), (\ref{eq:EbDyn}), (\ref{eq:Eblims}), (\ref{eq:powergrid}), (\ref{eq:periodicity}),(\ref{eq:scenario})\\
\end{aligned}
\end{equation*}
\end{prob}
where at  each $k \in \mathcal{K}$, we solve for the decision variable $\pmb{P}_{\mathrm{ib}, k} \in \mathbb{R}^{N_\mathrm{w} \times N_s}$.

For comparison, we benchmark the performance of the Stochastic and Recourse MPC described, with two alternative deterministic MPCs, one using perfect forecasts (Omniscient) and another one using point-estimate forecasts from~\eqref{eq:f_det} (Deterministic). The deterministic MPCs are defined by setting $N_s = 1$ in Problem~\eqref{prob:mpcSce}, since there is no measure of uncertainty for the predicted variables ($n_\alpha=0$).

\subsection{Simulated Environment}
For realistic evaluations, we use a closed-loop implementation with a simulated environment of the energy hub using observed values of $P_\mathrm{ev}$, $P_\mathrm{pv}$ and $p_\mathrm{el}$, and using the model defined by~\eqref{eq:Pbalance}, \eqref{eq:BlossOriginal}, \eqref{eq:Bloss2}, \eqref{eq:EbDyn}, \eqref{eq:powergridOriginal}, where there is no variable uncertainty ($N_s = 1$).
At each $k$, we solve Problem~\eqref{prob:mpcSce} to obtain the optimal control trajectory for each scenario $s$: $\pmb{P}_{\mathrm{ib}, k}^{\star s} = [P_{\mathrm{ib}, 0|k}^\star, ..., P_{\mathrm{ib}, N_\mathrm{w}-1|k}^{\star s}]^\top \in \mathbb{R}^{N_\mathrm{w}}$. Afterwards, we choose the first element of the control trajectory ($P_{\mathrm{ib}, 0|k}^\star$) as the action, which, neither for the Stochastic nor for the Recourse MPC, depends on $s$ due to~\eqref{eq:scenario}. Then we obtain $P_{\mathrm{b}, 0|k}^\star$ from~\eqref{eq:Bloss1} and apply it to the simulated environment. The environment returns the resulting observed $E_{\mathrm{b}, 1|k}$ and the observed $C_{\mathrm{el}, 0|k}$, based on the action applied ($P_{\mathrm{b}, 0|k}^\star$), the observed $P_{\mathrm{g}, 0|k}$ and the observed residual load ($P_{\mathrm{ev}, 0|k} - P_{\mathrm{pv}, 0|k}$). This process is repeated until $k=N_\mathrm{ep}-1$, when a new episode starts, resetting $k$. 

\section{Case Study Results}
We use publicly available data of $P_\mathrm{ev}$ \cite{GholizadehMusilek2024} and $P_\mathrm{pv}$ \cite{openaiAPI2024}, for which both sites are located in California, USA. For $p_\mathrm{el}$ and the $\mathrm{CO_2}$ emissions, we employ data for the Netherlands obtained from the European Network of Transmission System Operators for Electricity (ENTSO-E) \cite{entsoeAPI2024}. We use historical weather forecasts from the High-Resolution Rapid Refresh model by the National Oceanic and Atmospheric Administration of the US, available in the Open-Meteo.com Weather API \cite{openmeteoAPI2024}. More specifically, we work with an energy hub with a peak total EV power $\max(P_\mathrm{ev})=\unit[160]{kW}$ and a peak PV power $\max(P_\mathrm{pv})=\unit[70]{kW}$. For the BESS, we set the operational limits $\overline{E}=\unit[90]{kWh}$ and $\underline{E}=\unit[10]{kWh}$ and we initialize the battery as $E_{\mathrm{b},0} = \unit[25]{kWh}$. 
 We select 2021 data to test the accuracy and coverage of the forecasts and the performance of the controllers. For a fair comparison between different seasons, we design the test set by selecting 70 days per season from 2021. Each day selected is an evaluation episode with length $N_\mathrm{ep}=96$. More details on data handling are available in the public repository of the article.

\subsection{Forecasting Accuracy and Coverage}
We evaluate the accuracy of the forecasts with the Normalized Mean Absolute Error
\begin{equation}
    \mathrm{nMAE} = \frac{1}{N_\mathrm{o} \cdot(\mathrm{AE}_{\max} - \mathrm{AE}_{\min})}\cdot \sum_{i=1}^{N_\mathrm{o}}\mathrm{AE}_i,
\end{equation}
where $\mathrm{AE}_i = |Y_i-\hat{Y}_i|$, $Y_i$ is the observed value and $\hat{Y}_i$ is the point-estimate prediction, with the observation $ i \in \{ 1, ..., N_\mathrm{o}\}$, $N_\mathrm{o}$ is the number of observations in the test set, $\mathrm{AE}_{\max} = \max(\{ \mathrm{AE}_i \})$ and $\mathrm{AE}_{\min} = \min(\{ \mathrm{AE}_i \})$. Also, we evaluate the coverage of the prediction interval
\begin{equation}
    \mathrm{CPI}_\alpha = \frac{1}{N_\mathrm{o}}\cdot \sum_{i=1}^{N_\mathrm{o}} \mathds{1}_{\mathrm{in}, i}^\alpha,
\end{equation}
where $\mathds{1}_{\mathrm{in},i}^\alpha$ indicates if observation $i$ is within the prediction interval
\begin{equation}
    \mathds{1}_{\mathrm{in}, i}^\alpha = 
    \begin{cases}
        1 & \text{if } Y_i \in \mathrm{PI}_\alpha \\
        0& \text{otherwise}.\\
    \end{cases}
\end{equation}

\begin{figure}[!htb]
	\centering
	\includegraphics[width=0.80\linewidth]{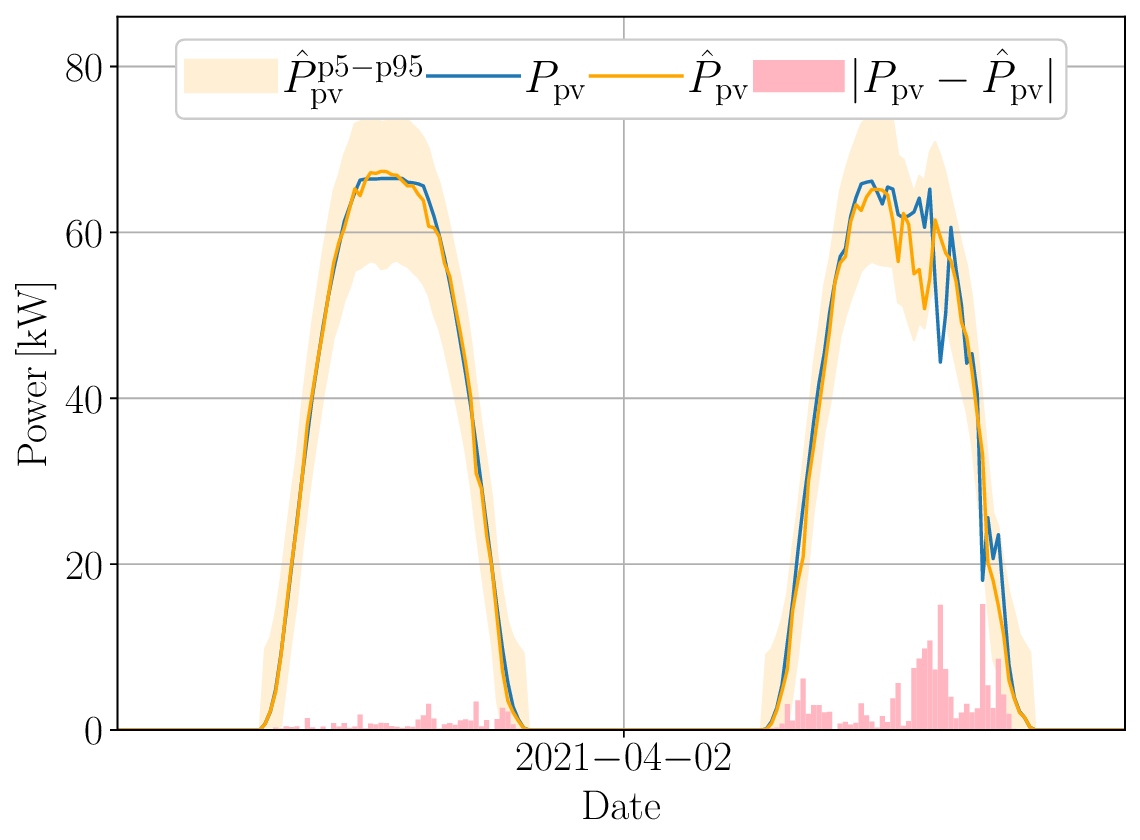}
	\caption{Observed $P_{\mathrm{pv}}$, \unit[24]{h} ahead prediction ($\hat{P}_{\mathrm{pv}}$) and prediction interval with $\alpha = 0.1$ ($\hat{P}_{\mathrm{pv}}^{\mathrm{p5-p95}}$) at hour 0, and absolute error ($|P_{\mathrm{pv}} - \hat{P}_{\mathrm{pv}}|$) for two days in spring.}
	\label{fig:pv}
\end{figure}

\begin{figure}[!htb]
	\centering
	\includegraphics[width=0.80\linewidth]{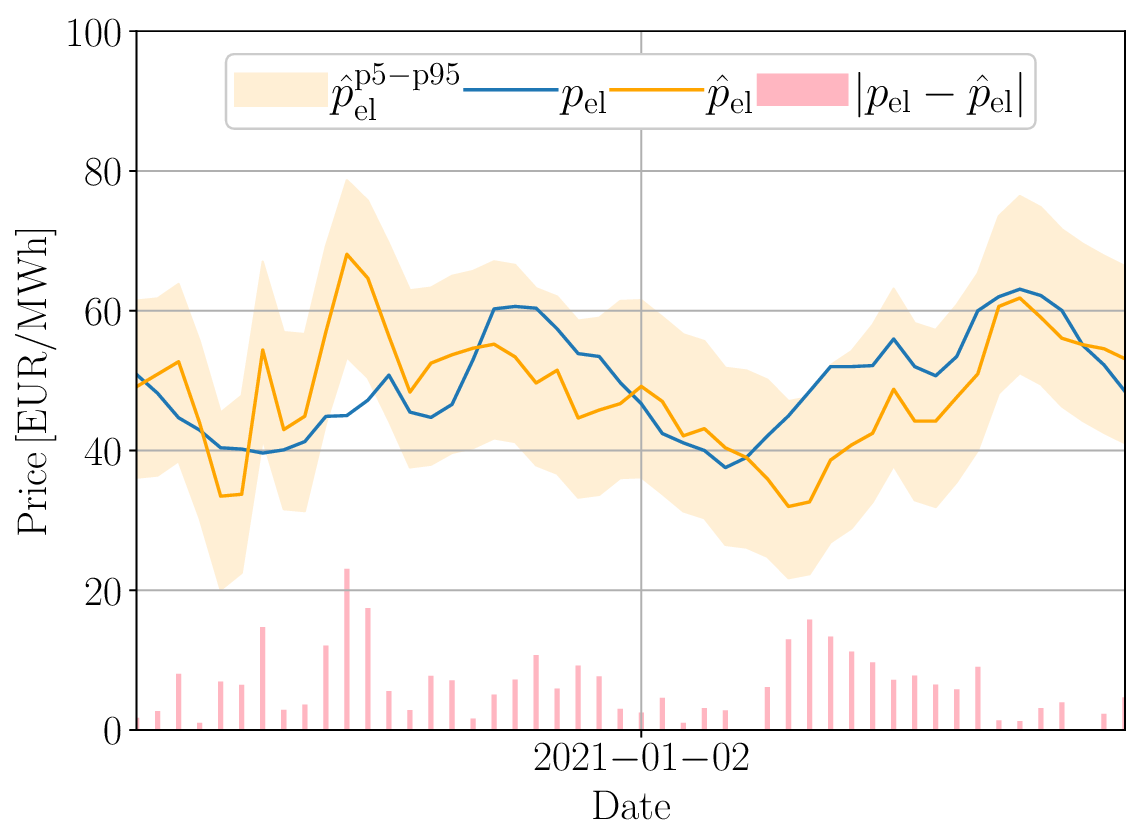}
	\caption{Observed $p_{\mathrm{el}}$, \unit[24]{h}-ahead prediction ($\hat{p}_{\mathrm{el}}$) and prediction interval with $\alpha = 0.1$ ($\hat{p}_{\mathrm{el}}^{\mathrm{p5-p95}}$) at hour 0, and absolute error ($|p_{\mathrm{el}} - \hat{p}_{\mathrm{el}}|$) for two days in winter.}
	\label{fig:da}
\end{figure}

\begin{figure}[!htb]
	\centering
	\includegraphics[width=0.83\linewidth]{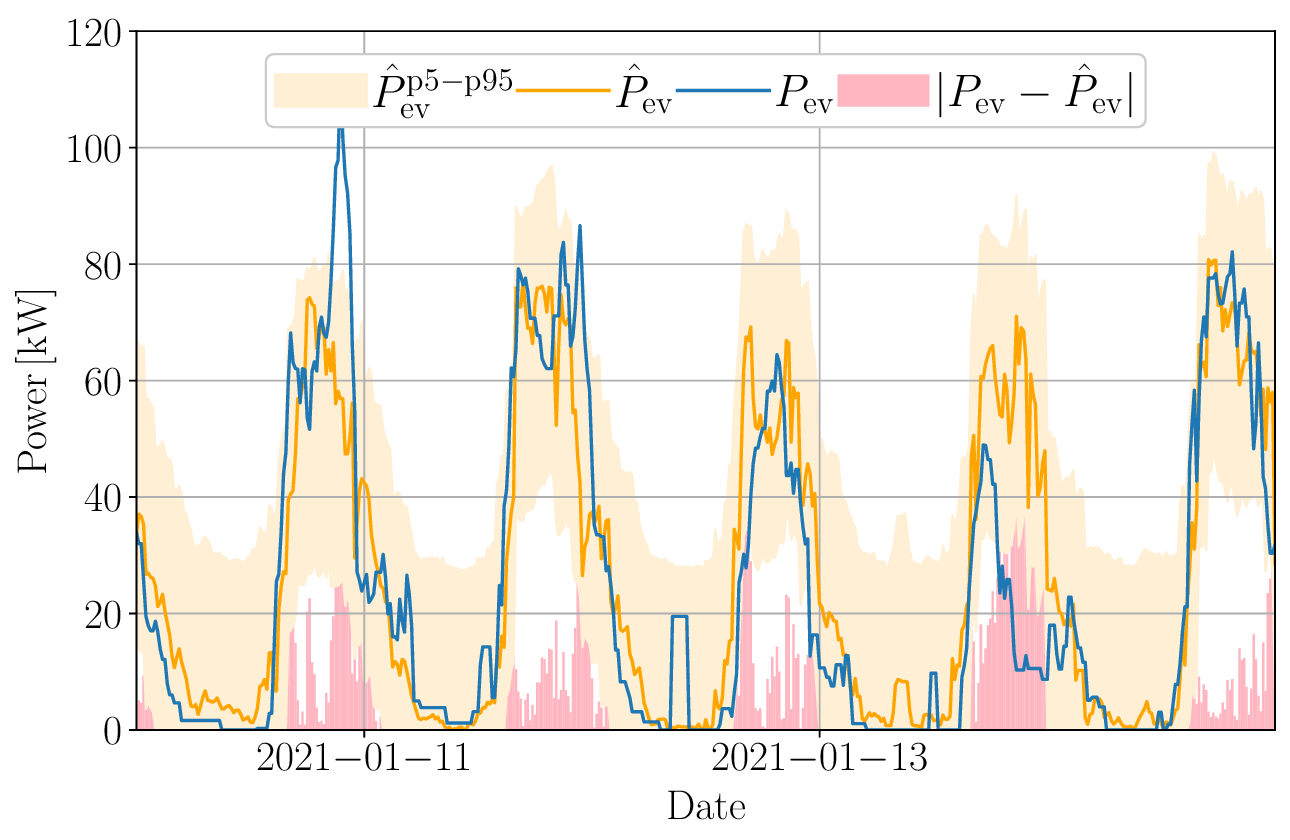}
	\caption{Observed $P_{\mathrm{ev}}$, \unit[24]{h}-ahead prediction ($\hat{P}_{\mathrm{ev}}$) and prediction interval with $\alpha = 0.1$ ($\hat{P}_{\mathrm{ev}}^{\mathrm{p5-p95}}$) at hour 0, and absolute error ($|P_{\mathrm{ev}} - \hat{P}_{\mathrm{ev}}|$) for five days in winter.}
	\label{fig:ev}
\end{figure}

When forecasting $P_\mathrm{pv}$, there is a seasonality of the forecasting accuracy due to changing weather conditions. In Tab. \ref{tab:pred_nmae}, we see that for summer and spring days, which have more stable weather conditions in this case, the predictions are more accurate. In Fig. \ref{fig:pv}, we show an example of a \unit[24]{h}-ahead prediction of $P_\mathrm{pv}$ for two days in spring. For $p_{\mathrm{el}}$, we see a degradation of the predicting accuracy as 2021 progresses, with the largest $\mathrm{nMAE}$ in the Autumn. We explain this by a regime change in the day-ahead electricity market, driven by the increasing gas prices in 2021. This highlights the additional difficulty of forecasting $p_\mathrm{el}$ and points toward a model improvement by introducing additional lagged features of the price or adding a feature to explain this regime change, e.g., the daily price of gas. Furthermore, in Tab. \ref{tab:pred_pi} we see that the prediction interval of $p_\mathrm{el}$ also degrades towards the end of 2021 and does not correspond to $\alpha = 0.1$, as opposed to $\mathrm{CPI}_{\mathrm{ev}, 0.1}$ and $\mathrm{CPI}_{\mathrm{pv}, 0.1}$. In Fig. \ref{fig:da}, we show an example of a \unit[24]{h}-ahead prediction of $p_\mathrm{el}$ for two days in winter. We compute \unit[24]{h}-ahead predictions of $P_{\mathrm{ev}}$ recursively, since we use the lagged intraday power feature defined in~\eqref{eq:featEvlagIntra}. For this reason, the forecasting accuracy of $P_\mathrm{ev}$ improves for shorter prediction horizons. In Fig. \ref{fig:ev}, we show an example of a \unit[24]{h}-ahead prediction of $P_\mathrm{ev}$ for two days in spring.
\vspace{5pt}
\begin{table}[H]
	\centering
	\caption{Normalized Mean Absolute Error of the \unit[24]{h}-ahead prediction at hour 0 per season for $P_\mathrm{ev}$, $P_\mathrm{pv}$ and $p_\mathrm{el}$ during a 280-day test.}
	\label{tab:pred_nmae}
	\begin{tabular}{c|c|c|c}
		Season & $\mathrm{nMAE}_\mathrm{ev}$ & $\mathrm{nMAE}_\mathrm{pv}$ & $\mathrm{nMAE}_\mathrm{el}$ \\ \hline
		Winter	& 0.107	&0.064	&0.039	\\
		Spring	&0.109	& 0.049&0.059 \\
		Summer	& 0.101	&0.054	&0.067	\\
		Autumn	& 0.109	&0.062	& 0.228 \\ \hline
		All & 0.106 & 0.057 & 0.098 \\
	\end{tabular}
\end{table}
\begin{table}[H]
	\centering
	\caption{Coverage of the \unit[24]{h}-ahead prediction interval with $\alpha=0.1$ at hour 0  per season for $P_\mathrm{ev}$, $P_\mathrm{pv}$ and $p_\mathrm{el}$ during a 280-day test.}
	\label{tab:pred_pi}
	\begin{tabular}{c|c|c|c}
		Season & $\mathrm{CPI}_{\mathrm{ev},0.1}$ & $\mathrm{CPI}_{\mathrm{pv},0.1}$ & $\mathrm{CPI}_{\mathrm{el},0.1}$ \\ \hline
		Winter	& 0.89	&0.88	&0.83\\
		Spring	&0.90	& 0.95&0.69\\
		Summer	& 0.91	&0.94	&0.66\\
		Autumn	& 0.89	&0.87	&0.22 \\ \hline
		All & 0.90 & 0.91 &  0.60 \\
	\end{tabular}
\end{table}
\vspace{3pt}
\subsection{Control Performance}
\begin{figure*}[!htb]
	\centering
	\includegraphics[width=0.92\textwidth]{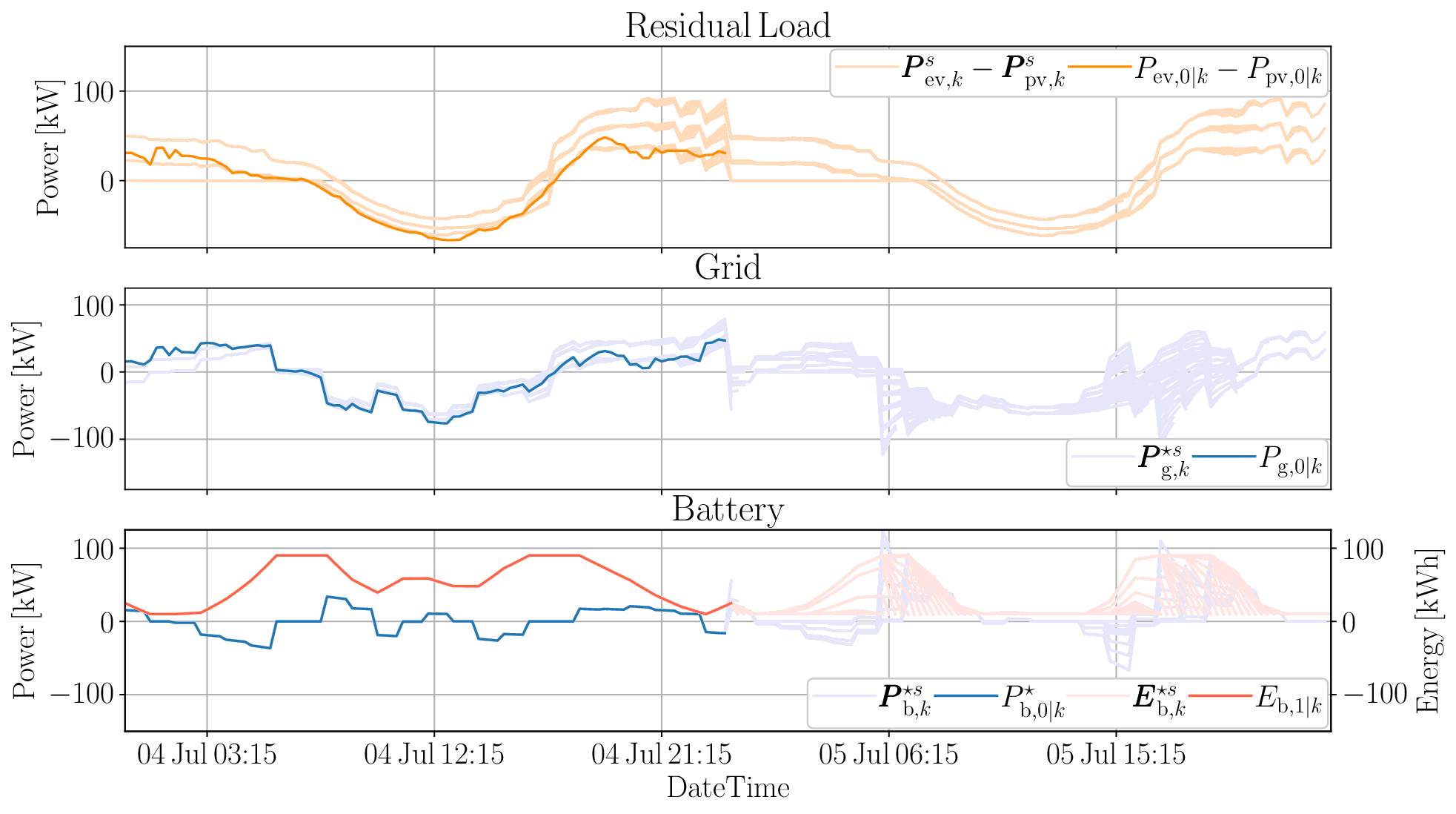}
	\caption{Energy hub operation during a one-day evaluation using the Stochastic MPC with a 96-step horizon window with a step resolution $\Delta T = \unit[0.25]{h}$. At each $k$, we present all scenarios obtained from the probabilistic \unit[24]{h}-ahead predictions of $P_\mathrm{ev}$ and $P_\mathrm{pv}$, as the residual load ($\pmb{P}_{\mathrm{ev}, k}^s - \pmb{P}_{\mathrm{pv}, k}^s$) along with the observed residual load ($P_{\mathrm{ev}, 0|k} - P_{\mathrm{pv}, 0|k}$). We include the optimal trajectories obtained from the Stochastic MPC at each $k$ ($\pmb{P}_{\mathrm{b}, k}^{\star s}$, $\pmb{P}_{\mathrm{g}, k}^{\star s}$ and $\pmb{E}_{\mathrm{b}, k}^{\star s}$ ). We also present, at each $k$, the observed $P_{\mathrm{g}, 0|k}$ and $E_{\mathrm{b}, 1|k}$ computed in the simulated environment of the energy hub, based on the applied action ($P_{\mathrm{b}, 0|k}^\star$) and $P_{\mathrm{ev}, 0|k} - P_{\mathrm{pv}, 0|k}$. }
	\label{fig:main}
\end{figure*}

\vspace{3pt}
In Fig. \ref{fig:main}, we present the result of a one-day evaluation using the Stochastic MPC defined in Problem (\ref{prob:mpcSce}). Then, we compare the Stochastic MPC with the Recourse, the Deterministic and the Omniscient MPC during 280 days of evaluation. In Tab. \ref{tab:control_cost} and \ref{tab:control_emission} we report the results using as metrics the observed daily net cost of operation and the daily net $\mathrm{CO_2}$ emissions of the grid, respectively. The Stochastic and Recourse MPC decrease the average cost of operation by 0.9\% and the average $\mathrm{CO_2}$ emissions by 0.3\%, when compared to the Deterministic Version. This also highlights that, in this case, using a point-estimate prediction  with no measure of uncertainty (Deterministic MPC) yields acceptable performance. The slight performance improvement after accounting for variable uncertainties, does not come at the cost of critical computational time. In Tab. \ref{tab:controltime}, we show that Stochastic and Recourse MPC do not increase episodic computational time over \unit[0.75]{min}, leaving enough time to compute optimal actions at every step $k$, by remaining far below $\Delta T$.
\vspace{8pt}
\begin{table}[H]
\centering
\caption{Normalized average daily observed $C_\mathrm{el}$ [\%] per season and per MPC version during a 280-day test. We normalized the values with respect to the Omniscient MPC.}
\label{tab:control_cost}
\begin{tabular}{c|c|c|c|c}
Season & Omniscient & Deterministic & Stochastic & Recourse \\ \hline
Winter & 100 & 108.92 &	108.03	& 108.04\\
Spring & 100&	115.29	& 114.99&115.0\\
Summer & 100&	109.29&	108.51&	108.5\\
Autumn&	100&	116.41&	115.37&	115.37\\ \hline
All & 100 & 113.65 & 112.76 & 112.77 \\
\end{tabular}
\end{table}
\begin{table}[H]
	\centering
	\caption{Normalized average daily observed $\mathrm{CO_2}$ emissions [\%] per season and per MPC version during a 280-day test. We normalized the values with respect to the Omniscient MPC.}
	\label{tab:control_emission}
	\begin{tabular}{c|c|c|c|c}
		Season & Omniscient & Deterministic & Stochastic & Recourse \\ \hline
		Winter & 100 & 100.98 &	100.73	& 100.73\\
		Spring & 100&	102.3	& 102.1&102.08\\
		Summer & 100&	100.72&	100.33&	100.33\\
		Autumn&	100&	101.75&	101.54&	101.54\\ \hline
		All & 100 & 101.4 & 101.13 & 101.13 \\
	\end{tabular}
\end{table}
\begin{table}[H]
\centering
\caption{Average runtime per episode [\unit{min}] and normalized runtime [\%] per MPC version with respect to Stochastic MPC during a 280-day evaluation.}
\label{tab:controltime}
\begin{tabular}{c|c|c}
MPC verison & Average Time & Normalized Time\\ \hline
Omniscient & 0.043 & 5.80\\
Deterministic & 0.321& 43.3\\
Stochastic & 0.743& 100\\
Recourse&	0.604& 81.3\\
\end{tabular}
\end{table}

\section{Conclusion}
In this work, we proposed two scenario-based model predictive controls that use probabilistic day-ahead forecasts for cost-optimal operation of an energy hub with solar generation, electric vehicle charging demand, a battery energy storage system and a grid connection. We also leveraged conformal prediction for calibrated distribution-free uncertainty quantification of solar generation, charging demand and day-ahead electricity prices. During a 280-day evaluation of the controllers in closed-loop interaction with a simulated environment, we reported a slight performance improvement of the Stochastic MPC and the Recourse MPC, compared to the Deterministic MPC, for all seasons of the evaluation year. Finally, we highlighted that point-estimate predictions yield acceptable performance and result in a deterministic controller with lower computational time than the two scenario-based alternatives. However, the computation time of each decision step with the Stochastic MPC and the Recourse MPC was far below the time resolution of the discretization. In conclusion, the Stochastic MPC and the Recourse MPC were the best overall performing controllers for this case study. Future work will explore jointly optimizing charging power levels and schedules of individual vehicles.
\newpage
\section*{Acknowledgment}
We thank Dr.\ I.\ New, Finn Vehlhaber and Juan Pablo Bertucci for proofreading this paper. This work was supported by RVO within the Charging Energy Hubs project with number NGFS23020 and was partially funded by the Dutch National Growth Fund.
\vspace{-10pt}
\bibliographystyle{IEEEtran}
 \bibliography{../../../Bibliography/energyhubs.bib}
 
\end{document}